\documentclass[12pt]{article}
\usepackage{graphicx}
\usepackage{epsfig}
\usepackage{epsf}
\usepackage{amssymb}
\usepackage{rotate}
\newcommand{\ba}{\begin{array}}
\newcommand{\ea}{\end{array}}
\newcommand{\bd}{\begin{displaymath}}
\newcommand{\ed}{\end{displaymath}}
\newcommand{\be}{\begin{equation}}
\newcommand{\ee}{\end{equation}}
\newcommand{\bea}{\begin{eqnarray}}
\newcommand{\eea}{\end{eqnarray}}

\def\a{\alpha}
\def\b{\beta}

\def\e{\epsilon}

\def\etal {{\em et al.,\ }}
\def\th13 {\theta_{13}}

\begin{document}
\thispagestyle{empty}
\begin{flushright}
\texttt{CU-PHYSICS/04-2007}\\

\end{flushright}
\vskip 15pt

\begin{center}
{\Large {\bf The neutrino mass scale and  the mixing 
angle $\theta_{13}$ for quasi-degenerate Majorana neutrinos 
}}
\renewcommand{\thefootnote}{\alph{footnote}}

\hspace*{\fill}

\hspace*{\fill}

\large{\bf{
Rathin Adhikari $\footnote{E-mail address:
rathin@jamia-physics.net}^{\dagger}$,
 Anindya Datta$\footnote{
E-mail address: adphys@caluniv.ac.in}^{\star \ddagger}$
}}\\
\large{\bf{ and Biswarup Mukhopadhyaya
$\footnote{E-mail address: biswarup@mri.ernet.in}^{\ddagger}$}}
\end{center}
\begin{center}
\small$\phantom{i}^{\dagger}${\em Centre for Theoretical Physics,\\ Jamia Millia
Islamia, New Delhi - 110025, India}\\

\vskip 10pt

\small$\phantom{i}^{\star}${\em Department of Physics, University of Calcutta,\\
92, A. P. C. Road, Kolkata 700009, India }
\\
\vskip 10pt

\small$\phantom{i}^{\ddagger}${\em Harish-Chandra Research Institute,\\
Chhatnag Road, Jhunsi, Allahabad - 211 019, India }

\vskip 40pt

{\bf ABSTRACT}

\vskip 0.5cm

\end{center}

Considering a general mass matrix for quasi-degenerate neutrinos and
treating the experimentally known oscillation parameters as inputs, we
study the correlation between the degenerate mass scale $(m)$ and the
mixing angle $\theta_{13}$. We find that, corresponding to different
values of $m$, there exist upper bounds on $\theta_{13}$, so that a
precise determination of the latter in future may put upper limit on
the former, and vice versa.
  One can also find a possible
correlation between $m$ and lower bound of $\theta_{13}$, depending on
the relative strength of the unperturbed degenerate mass matrix and
the perturbation.
The possible constraints on the parameters
of  few models of quasi-degenerate neutrinos are briefly  discussed.
  

\newpage
    
The currently available experimental data on neutrino oscillation
indicate that the mass squared differences of neutrinos are quite
small.  Although the mass-squared differences and mixing angles
(except $\theta_{13}$) are known from neutrino oscillation
experiments, the absolute values of the neutrino masses still remain
unknown. Depending on the scale, the pattern of neutrino masses may be
hierarchical or quasi-degenerate in nature.  In particular, if one
assume that relic neutrinos constitute the hot dark matter of the
universe, then the scale of neutrino masses are expected to be
somewhat higher than their differences, and neutrinos are expected to
be quasi-degenerate. Considering the recent results \cite{wmap} from
WMAP experiment, there might be some improvements in the upper bound
on the neutrino mass.  However, keeping in mind all possible uncertainties
in the cosmological bounds, an upper limit of 2 eV on the sum of the masses 
of three neutrino flavours can be said to exist \cite{fukugita}.


Quasi-degeneracy of neutrinos has also been suggested in a number of
theoretical models proposed in the literature
\cite{deg1}.  For example, considering neutrino masses as degenerate
at some seesaw scale, various authors \cite{ramond,sher, rn, adhi, ma,
rn1} have shown that large mixing angles for solar as well as
atmospheric neutrinos can be obtained after extrapolating masses and
mixing to the weak scale using renormalization group equations.

Among the neutrino oscillation parameters, the values of two
mass-squared differences and two mixing angles, viz.  $\theta_{12}$
and $\theta_{23}$, are already known to a reasonable degree from the
solar \cite{solar} and atmospheric \cite{atm} neutrino data.  However,
quantities still unknown are the mixing angle $\theta_{13}$, mass
scale $m$ of the neutrinos and the $CP$ violating phase $\delta$ in
the neutrino mixing matrix.  It should be noted that the
CHOOZ-Palo-Verde experiments \cite{reac, theta13} have put an upper
bound on $\theta_{13} (< 12^\circ)$. Furthermore, the absence of
neutrino-less double beta decay implies an upper bound on $m$ \cite{nuless2beta},
which is compatible with the limit coming from the hot dark matter
content of the universe \cite{wmap}.

In this note, we would like to illustrate the interplay among the unknown
quantities, namely, $m$, elements of the perturbation matrix and
$\theta_{13}$, assuming that neutrinos are of Majorana nature and the
masses are quasi-degenerate. We have taken into account the solar and
the atmospheric data as well as neutrino less double beta decay
constraint in our analysis, which has not been considered in earlier
analyses.  As we shall see, the maximum allowed value of $\theta_{13}$
and the cosmological upper limit on the overall neutrino mass scale
actually restrict the parameter space of the perturbation matrix
lifting the degeneracy. This can enable one to constrain various
theoretical models of the neutrino mass matrix, which in turn
determine the way degeneracy is lifted. We will also try to find any
possible correlation between the degenerate mass scale $m$ with
minimum value of $ \theta_{13}$.

Without going to a specific model, we shall first consider the most
general form of degenerate mass matrix in the weak interaction basis,
which allows mixing among different flavours of neutrinos due to
different intrinsic $CP$-properties. This degeneracy is lifted, for
example, by the breaking of some symmetry at the seesaw scale \cite{RT_new}.
Thus we add a small perturbation matrix to the original degenerate
mass matrix in a model-independent way.  We use as our inputs the
known oscillation parameters, namely:

\begin{itemize}
\item $|\Delta m^{2}_{23}|\simeq 2.12^{+1.09}_{-0.81}\times 10^{-3}$
eV$^2$, $\theta_{23}\simeq$ ${45.0^\circ}^{+10.55^\circ}_{-9.33^\circ}$
(from the atmospheric $\nu_\mu$ deficit \cite{atm}).

\item $\Delta m^{2}_{12} \simeq 7.9^{+1.0}_{-0.8} \times 10^{-5}$ eV$^2$,
$\theta_{12}\simeq$ ${33.21^\circ }^{+4.85^\circ}_{-4.55^\circ}$
(from the solar $\nu_e$ deficit \cite{solar}). 
\end{itemize}
\noindent
where   $\Delta m^{2}_{ij}$= $m^{2}_{j} -
m^{2}_{i}$. 

Thereafter, the application of degenerate perturbation theory, in
conjunction with the constraints arising on the perturbation matrix
after using the above experimental values, allows us to obtain the
correlation sought after.  Specific choices of models for generating
neutrino mass are likely to restrict further the general form of the
perturbation matrix. We study the correlation in the context of a few
models having family symmetries of abelian and non-abelian nature, and
check whether these in turn can constrain some parameters of the
models even further.
 
With the charged lepton matrix taken as diagonal, real and positive,  
the Majorana mass terms in the flavour basis can be expressed as 
\be
{\cal L}_{mass} = - {\left( \nu_{L_\a}\right)}^T C^{-1} M_{\a \b} \nu_{L_\b} 
+ h.c
\ee
where $M_{\a \b}$ is a $ 3 \times 3$ complex symmetric mass matrix and
$\nu_{L \a}$ is the weak eigenstate basis of neutrinos corresponding to 
three generations. 
The neutrino flavour states $|\nu_\alpha
\rangle$, $\alpha = e, \mu, \tau$, in the weak basis are related to
the neutrino mass eigenstates $|\nu_i\rangle$, $i=1,2,3$, with
masses $m_i$ :
\begin{equation}
\vert\nu_\alpha \rangle = \sum_i U_{\alpha i} \vert\nu_i\rangle~.
\end{equation}
where $U$ is a  $3 \times 3$ unitary matrix. In general, the 
mass matrix $M$ can
be diagonalised by a transformation of the form
\bea
U^T \;M \; U = M_{diag}  = \left( \begin{array}{ccc}
m_1  & 0 & 0 \\
0 & m_2 & 0 \\
0 & 0 & m_3 
\end{array} \right)  
\eea 
where $M_{diag}$ is the diagonal mass matrix. If
one considers the three masses to be degenerate, then $U$ can be
rotated away for Dirac neutrinos. This cannot, however, be done in the case of Majorana
neutrinos if the $CP$ property of one of the fields is different from
those of the other two.  It was shown by Branco {\it et al}
\cite{branco} that, in such a case, the fact that the $CP$-eigenvalue of
a Majorana state can be $+i$ or $-i$ implies a still non-trivial form
of the the diagonalising matrix $U$ which contains two mixing angles
and one phase. Implications of such (or similar) scenarios have also
been explored, for example, in reference \cite{ramond}.

For degenerate Majorana neutrinos  remembering that the mixing angles $\theta_{12}$ and 
$\theta_{23}$ are required to be large by observation, we shall consider
the intrinsic $CP$-eigenvalue associated with $\nu_2$ to be opposite to 
that of the
other two neutrinos. Without losing any generality we can thus write
\bea
U = \left( \begin{array}{ccc}
1 & 0 & 0 \\
0 & c_{23}^{\prime} & s_{23}^{\prime}  \\
0 & s_{23}^{\prime} & -c_{23}^{\prime} \\
\end{array} \right)  
\left( \begin{array}{ccc}
c_{12}^{\prime} & s_{12}^{\prime} & 0 \\
s_{12}^{\prime} & -c_{12}^{\prime} & 0    \\
0 & 0 & e^{-i \a} \\
\end{array} \right)
\left( \begin{array}{ccc}
1 & 0 & 0 \\
0 & e^{i \b} & 0  \\
0 & 0 & 1 \\
\end{array} \right)
\label{umat}
\eea where $c_{ij}^{\prime}=\cos\theta_{ij}^{\prime}$,
$s_{ij}^{\prime} =\sin\theta_{ij}^{\prime}$ and $\b = \pi/2$
corresponds to different intrinsic $CP$ property of $\nu_2$ with
respect to other two neutrinos and $\a$ corresponds to $CP$ violating
phase in general. We shall choose it to be zero for the $CP$
conserving case.   Following
ref. \cite{branco} the degenerate mass matrix $M$ in the flavour basis
can be written as 
\be M = m\; U^*  U^\dagger \ee 
where degenerate mass scale  $m = m_1 = m_2 = m_3$.

However, in spite of such a mixing, neutrinos cannot oscillate, since
the quantity governing oscillations are the mass-squared
differences. In order to lift the degeneracy, we assume that
degenerate mass scale $m$ is somewhat higher than the required
mass-squared differences as indicated by the oscillation
experiments. There are  studies on the mechanism for lifting
such degeneracy. One can assume degeneracy at a high scale and
envision that it is lifted through running, as some symmetry which
holds at high scale is broken. Alternatively, one can remain confined
to the electroweak scale itself and consider mass splitting effects
there. In either case, the mechanism consists in a perturbation to the
mass matrix, in a basis where it is diagonal and degenerate. This
perturbation matrix is real and symmetric in several studies in recent
past. We follow the same practice here. Moreover, most of these
investigations do not throw much light on the {\em numerical values} of the
elements of the perturbation matrix. This is the point we wish to address,
namely, how the limits on $\theta_{13}$ as well as the neutrino mass scale 
restrict the values of the perturbation matrix elements. We believe that 
, in spite of a six-fold multiplicity of the elements, such constraints can be 
useful in shortlisting viable models.

We thus consider a small perturbation to the degenerate
mass matrix $M_{diag}$, parameterized as

\bea Q = \e \left(
\begin{array}{ccc} e & a & b \\ a & g & 1 \\ b & 1 & f \\
\end{array} \right)
\label{pert_mat}
\eea
where $\e^2$ is of the order of $10^{-3}$ eV$^2$ and other parameters
are $\le 1$. In general, the smallness of $\theta_{13}$ present in
the standard parametrisation of neutrino mixing matrix (as mentioned in equation
 (15) below) hints at the
$CP$ violating effect being small in the neutrino sector. So keeping
$CP$ violating phase only in $M$ we have neglected it in the 
small perturbation matrix $Q$  making it real. 
 $Q$ is
written in a basis in which the degenerate mass matrix is diagonal.  It
is also obvious that in the flavour basis $Q$ becomes
\be
Q_0 = U^\ast \; Q \; U^\dagger
\ee
where  $U$ is of the form indicated in (4). Unlike $Q$ in the flavor basis $Q_0$ is
complex. 
We next employ the methodology of degenerate
perturbation theory, with the diagonal and degenerate mass matrix
determining the unperturbed basis. 
Besides, we absorb the phase associated with $\b = \pi/2$ 
in the Majorana neutrino field, thus enabling it to disappear
from the matrix $U$ and we define this as $U_1$.  Considering $M$ in (5)
and using   $U_1$ as the diagonalising matrix
from  eq. (3) we get the diagonal elements of $M_{diag}$ as , 
\be
m_2 = -m, \; m_1 = m_3 = m
\label{masses} 
\ee
Now, only  $m_1$ and $m_3$ are degenerate. 
If we wish to parameterize the lifting of degeneracy in such
a manner that the mass $m_1$ remains unchanged, then one
may set the following condition on the parameters in $Q$:
\be
b^2 = e f 
\label{b2ef}
\ee
Using first order degenerate perturbation theory, 
we obtain the following mass eigenvalues  for $M_{diag} + Q$, lifting the 
degeneracy to first order in $\epsilon$:
\be
m_1 = m ,\; m_2 = -m + \e g ,\; m_3 = m + \e (e +f)
\label{pert_mas}
\ee
and the diagonalising matrix as :
\bea
U^{\prime} = \left( \begin{array}{ccc}
\cos \psi  &  
- a r &   
\sin \psi   \\
r \left(a \cos \psi - \sin \psi \right) &
 1 &
 r \left(a \sin \psi + \cos \psi \right) \\
 -\sin \psi & -r  & \cos \psi   \\
\end{array} \right)\;{\cal N} 
\eea  
where  $\tan 2 \psi = 2 b /(f-e)$; $r = {\epsilon \over {2m}}$. 
${\cal N} \equiv diag(N_1 ^{-1}, N_2 ^{-1}
,N_3^{-1})$ where $N_1, N_2, N_3$ are the proper normalisation constants 
for the eigenvectors. These can be easily expressed in terms of the 
parameters of the theory.

\bea
{\cal N}_1 &=& \left[1 + r^2 (a \cos \psi - \sin \psi)^2 \right]^
{\frac{1}{2}} \nonumber \\
{\cal N}_2 &=& \left[ 1 + r^2 (1 + a^2) \right]^
{\frac{1}{2}} \nonumber \\
{\cal N}_3 &=& \left[1 + r^2 (a \sin \psi + \cos \psi )^2 \right]^
{\frac{1}{2}} \nonumber\\
\eea

To go to the physical mass eigenstate basis, however,
we keep $\b = \pi/2$ in $U$ 
and now the eigenvalues are same as in equation \ref{masses}
except there will be overall change in sign in $m_2$ making it positive. 
So replacing $U_1$ by $U$ we write
\be
U_0 = U U^{\prime} 
\label{uup}
\ee
which will diagonalise the mass matrix $M+Q_0$. Although there
was no mixing of $\nu_e$ and $\nu_\tau$ in $U$ but $U_0$ has that
mixing.   
One may note that, since the solar neutrino data indicates that $m_2$
is heavier than $m_1$ and that there is no sign ambiguity in $\Delta
m_{12}^2$, the parameter $g$ should be negative in our case. To relate the
different parameters with the experimental data we shall write $U_0$
in the standard form of a $3\times 3$ unitary mixing matrix:
\bea
U_0 = \left(
          \begin{array}{ccc}
          c_{12}c_{13} & s_{12}c_{13} & s_{13}e^{-i\delta}  \\
 -c_{23}s_{12} - s_{23}s_{13}c_{12}e^{i\delta} & c_{23}c_{12} -
s_{23}s_{13}s_{12}e^{i\delta}&  s_{23}c_{13}\\
  s_{23}s_{12} - c_{23}s_{13}c_{12}e^{i\delta}& -s_{23}c_{12} -
c_{23}s_{13}s_{12}e^{i\delta} & c_{23}c_{13} \end{array} \right),
\label{standerd_para}
\eea
where $c_{ij}=\cos\theta_{ij}$, $s_{ij}=\sin\theta_{ij}$ and $\delta$
is the $CP$ violating phase.

  The parameters of $Q$ in equation \ref{pert_mat} can be constrained
from the mass squared differences obtained from solar and atmospheric
neutrino data. Using equation \ref{pert_mas} up to the order of $\e$, one can
write
\be
2m \e g = \Delta m^{2}_{12}   
, \;\; 
2m \e (e + f) = \Delta m^{2}_{23} 
\ee
From equation \ref{b2ef} it follows that $e$ and $f$ cannot be of opposite sign 
which implies
\be
  Q_{11} = \e e \leq \Delta m^{2}_{23}/2m
\ee

It follows from the above that 
\be 
x (\equiv e m^2)  \leq \Delta m^{2}_{23}|/r
\label{xlim}
\ee

We consider $\e^2 a^2  < |\Delta m^{2}_{12}|$ so that  
solar neutrino data is satisfied.

We further simplify the analysis by setting $\delta =0$ in
equation \ref{standerd_para} and $\a =0 $ in equation \ref{umat} which correspond to $CP$
conserving case.  Comparing 11, 12 and 13 elements of $U_0$ in eqs. \ref{uup}
and \ref{standerd_para}, we obtain
\be
\sin \theta_{13} = N_3^{-1}\;\left[ \cos \theta_{12}^{\prime} \sin \psi + 
\sin \theta_{12}^{\prime} \left( a \sin \psi + \cos \psi \right) 
{\e \over 2m } \right]
\label{thet13}
\ee
where

\be
\tan \theta_{12}^{\prime} = \left[ { a c_{12} \e + 2m s_{12}
\cos \psi \over 2m c_{12} - \e a s_{12} \cos \psi + \e s_{12} \sin \psi
} \right] 
\label{tan12}
\ee

Equations \ref{thet13} and \ref{tan12} can be re-expressed as in term of $x ~(\equiv
e m^2$):

\be
\sin \theta_{13} = {\cal N}_3^{-1} \left[ \cos \theta^{\prime}_{12}\; \sqrt{\frac{2 r x}{\Delta m^2_{23}}} + r \sin \theta^{\prime}_{12}\;(\sqrt{1 -\frac{2 r x}{\Delta m^2_{23}}} + 
a \sqrt{\frac{2 r x}{\Delta m^2_{23}}}) \right]
\label{theta13}  
\ee 

and 

\be
\tan \theta_{12}^{\prime} = \frac{\frac{{\cal N}_2}{{\cal N}_1} \tan \theta_{12} \sqrt{1 - 
\frac{2 r x}{\Delta m^2_{23}}} + a r}{1 - r\left( a \sqrt{1 - 
\frac{2 r x}{\Delta m^2_{23}}} - \sqrt{\frac{2 r x}{\Delta m^2_{23}}}\right)}
\label{theta12}
\ee

Considering $\theta_{12}$ from the solar neutrino experimental data as
discussed in the introduction, one can find the relation between the
two unknowns $m$ and $\theta_{13}$ in terms of the elements of $Q$.
It may be noted here that for both normal and inverted hierarchy
equations \ref{theta13} and \ref{theta12} remain unchanged.

Now one can see how $\theta_{13}$ and $m$ are related. It is evident
that for a given value of $r$, $\theta_{13}$ is a function of $x$ only. A
careful inspection of equation \ref{thet13}, tells the monotonically
increasing

\begin{figure}[htb]
\begin{center}
\hskip -0.0cm
\vskip -0.0cm
\psfig{figure=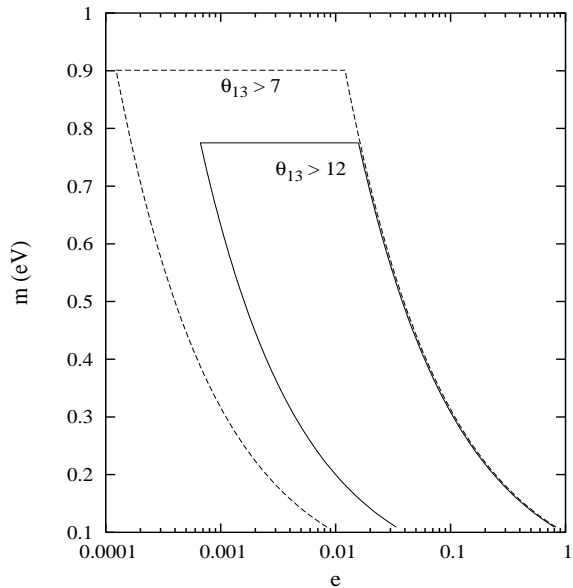,width=8.0cm,height=8.0cm,angle=0}
\caption{ \sf \small
Allowed region for $m$ and $e$ depending on $\theta_{13}$. $r \equiv
\frac{\e}{2m} = 0.1$ has been used in this plot. 
}
\label{allowed_region}
\end{center}
\end{figure}

\newpage

\noindent
dependence of $\theta_{13}$ on $x$. Thus maximum allowed
value of $\theta_{13}$ corresponds to the limiting value of $x$ as
defined in equation \ref{xlim}. The contours of $\sin \theta_{13}$ 
thus correspond to specific values of $x$. In other words, they corresponds to
contours of specific values of $x$ in the $e - m$ plane.

In figure \ref{allowed_region}, we vary $  e$ in the range
from 0 to its upper bound defined by equation
\ref{xlim}. Corresponding to different values of $\theta_{13}$ we have
shown the variation of $m$ as obtained numerically.  The region marked
with $\theta_{13} > 12$ is thus disallowed from the CHOOZ result.  We
have argued earlier how the normal and inverted hierarchical ordering
of neutrino masses would produce the same constant $\theta_{13}$
contours in the $m - e$ plane.  Depending on the possible values
of $e$, one can find an upper bound on $m$ in both normal and
inverted hierarchical case from this plot.  If, in future, better
bound on $\theta_{13}$ is obtained then it is possible to improve
upper bound on the degenerate mass scale, $m$, compared to its present
cosmological bound \cite{wmap}.

\begin{figure}[htb]
\begin{center}
\hskip -0.0cm
\psfig{figure=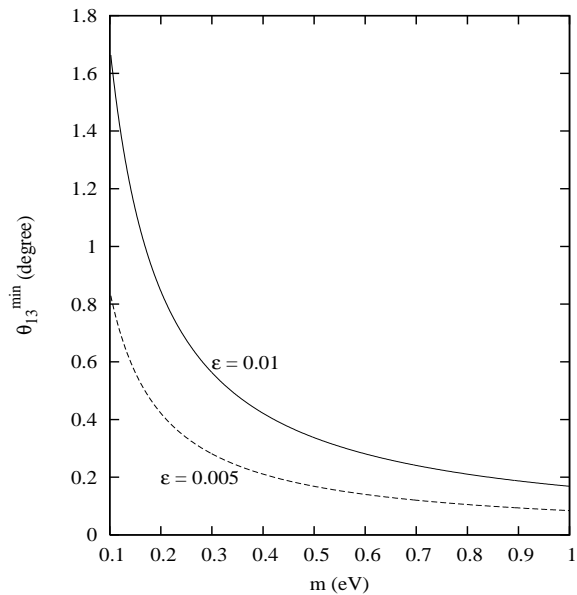,width=8.0cm,height=8.0cm,angle=0}
\caption{ \sf \small
Minimum value for $ \theta_{13}$  vs. the degenerate mass scale $m$ for two 
different values of perturbation parameter $\epsilon$. 
}
\label{min_theta_13}
\end{center}
\end{figure}

In figure \ref{min_theta_13}, we have plotted the minimum values for
 $\theta_{13}$ as a function of the degenerate mass scale $m$ for two different
values for the perturbation parameter $\epsilon$. The range over which we vary
$m$ is fixed by two considerations. The upper limit should be less than
1 eV  from
cosmological considerations \cite{wmap}. Furthermore, we would like to generate
the tiny mass square differences by perturbing the degenerate matrix (with
eigenvalues $m$). Thus the value of $m$ should be bigger than the differences (which
is generated by perturbation) in the masses. Thus, keeping in mind the
atmospheric mass (square) difference, we take the lower limit for $m$ to be
0.1 eV.
 
It is interesting to note that that there is lower bound on
$\theta_{13}$ depending on the value of $m$.  One can see that the minimum
 of $\theta_{13}$ decreases with
increasing  value of $m$. Increasing $m$ and keeping the value of $\epsilon$
fixed implies, the relative
strength of the perturbation with respect to the unperturbed degenerate
mass matrix,
keeps on decreasing. This in turn is taming down the value of
$ \theta_{13}$.
In the limit when perturbation vanishes, masses are 
exactly the same and $\theta_{13}$ is exactly zero. However, to
create qusi-degenerate masses there is perturbation and as such there
is some lower
bound on $\theta_{13}$ which will be nearer to  $1^0$ if the cosmological
bound on neutrino masses improves further.

The general degenerate mass matrix $M$ in the weak basis and $Q_0$
contain too many parameters. However, on the basis of some symmetry
principle one can reduce the number of these parameters. 
One may consider
a simple model of neutrino mass \cite{ma} using a leptonic
Higgs doublet and three right-handed singlet fermions at TeV energy scale 
or below.  Considering discrete $A_4$ symmetry one may obtain
the degenerate mass matrix of the form
\bea
M = m \left(  \begin{array}{ccc}
1 & 0 & 0  \\
0   & 0 & 1 \\ 
0  & 1 & 0 
\end{array}   \right)  
\eea
as discussed in  \cite{ma}. Such pattern correspond
to $\theta_{12}^{\prime} = 0$ and
$\theta_{23}^{\prime} = 3 \pi/4$ in the degenerate mass matrix $M$.
Then our perturbation matrix $Q_0$ in the flavor basis  is
\bea
Q_0= \epsilon \left( \begin{array}{ccc}
e &   (- i a +b)/\sqrt{2} &  ( i a +b )/\sqrt{2} \\
 (- i a +b)/\sqrt{2} & (-2i+f-g)/2 & (f+g)/2 \\
 (i a +b)/\sqrt{2} & (f+g)/2 & (2i+f-g)/2
\end{array} \right)
\eea
This can be obtained by considering soft symmetry-breaking
terms of the form $m_{ij} N_{iR} N_{jR}$ in the Lagrangian where $N$ are right-handed neutrino fields.
 Thus parameters $m_{ij} $ of the theoretical model
  of neutrinos will be related to the parameters of
  $Q_0$ and as such $Q$. From equation. \ref{theta13} it is evident that $\theta_{13}$ is very close
to $\psi \simeq \sqrt{\frac{2 m e \epsilon}{\Delta m^2_{23}}}$.
As  $\theta_{13} < 12^0 $ using (15) one gets the relative estimate of
the parameters $e$ and $f$. So $ f   \sim  2 \times 10^{-3} /(m \epsilon)$
and $e$ is much smaller than that as required by CHOOZ constraint.

Degenerate neutrino masses from an abelian family symmetry has
also been considered in \cite{ramond}.    The neutrino mass matrix as
considered in eq. (38) of that paper is similar in form as shown above
in (22) \& (23) provided that we consider  $ e=0 $  and $ f = - g $ in
our general form of the perturbation matrix  $Q$ so as to reproduce
the $Q_0$ appropriate for their mass matrix.

So far we have considered the case of different intrinsic $CP$ parities
of Majorana neutrinos for which there is mixing even in the exact
degenerate mass limit. However there may be a case for which intrinsic
$CP$ parities of all the Majorana neutrinos are the same \cite{sher}. 
In that case the $U$ matrix can be rotated away and the mixing will be controlled
by the perturbation matrix only. So it is not possible
to perform the same analysis
for the perturbed part in the quasi-degenerate mass matrix as done in this paper. 
Neutrino-less double beta decay in such a case may yield a 
direct constraint on the  approximate degenerate mass $m$, although
$\theta_{13}$ may not receive any additional constraint.
However, same $CP$ parity of all neutrinos  is disfavoured from the
viewpoint of the requirement of stability at the weak scale of
quasi-degenerate
neutrino masses and mixing
 pattern which emerge at the high scale  \cite{rn} as  the CHOOZ constraint
 on $\theta_{13}$ cannot be satisfied.   In ref \cite{rn1} although
  the same $CP$
   parity of quasi-degenerate neutrinos has been considered at
    the see-saw scale
   but at the weak scale it has been found that the appropriate mass-squared
   difference for oscillation of solar neutrinos
   cannot be obtained.

\newpage

\noindent
{\large{\bf {Acknowledgments}}}\\

We thank Amitava Raychaudhuri for helpful comments.
R. Adhikari acknowledges the hospitality
of Harish-Chandra Research Institute under the
DAE X-th plan project on collider physics 
while this work was in progress. A. Datta  acknowledges the hospitality
of Centre for Theoretical Physics, Jamia Millia Islamia, New Delhi.\\

\hspace*{\fill}


\begin{thebibliography}{99}
\bibitem{wmap}D. N. Spergel et al., astro-ph/0603449. 

\bibitem{fukugita}M. Fukugita et al.,  Phys.\ Rev.\ D {\bf 74}, 027302 (2006).

\bibitem{deg1} 
S. T. Petcov and A. Yu Smirnov, 
 Phys. Lett {\bf B 322}, 109 (1994);
D. O. Caldwel and R. N. Mohapatra,
Phys. Rev. {\bf D 48}, 3259 (1993), Phys. Rev. {\bf D 50}, 3477 (1994);  
  M.~Fukugita, M.~Tanimoto and T.~Yanagida,
  Phys.\ Rev.\ D {\bf 57}, 4429 (1998). 

\bibitem{ramond}
  P.~Binetruy, S.~Lavignac, S.~T.~Petcov and P.~Ramond,
  Nucl.\ Phys.\ B {\bf 496}, 3 (1997)

\bibitem{sher} 
C.~D.~Carone and M.~Sher,
  Phys.\ Lett.\ B {\bf 420}, 83 (1998)

\bibitem{rn}
  K.~R.~S.~Balaji, A.~S.~Dighe, R.~N.~Mohapatra and M.~K.~Parida,
  Phys.\ Rev.\ Lett.\  {\bf 84}, 5034 (2000).


\bibitem{adhi}  R.~Adhikari, E.~Ma and G.~Rajasekaran,
  Phys.\ Lett.\ B {\bf 486}, 134 (2000).

\bibitem{ma}
 E.~Ma and G.~Rajasekaran,
  Phys.\ Rev.\ D {\bf 64}, 113012 (2001);
  E. Ma, Phys. Rev. Lett. {\bf 86}, 2502 (2001).


\bibitem{rn1}
  R.~N.~Mohapatra, M.~K.~Parida and G.~Rajasekaran,
  Phys.\ Rev.\ D {\bf 69}, 053007 (2004); Phys.\
  Rev.\ D {bf 71}, 057301 (2005).

\bibitem{solar}
Y. Fukuda {\it et al.} (Super-Kamiokande Collaboration),
Phys. Rev. Lett. {\bf 81}, 1158 (1998);
Phys. Rev. Lett. {\bf 82}, 1810 (1999);
Phys. Rev. Lett. {\bf 82}, 2430 (1999);
S. Fukuda {\it et al.} (Super-Kamiokande Collaboration),
Phys. Rev. Lett. {\bf 86}, 5651 (2001);
Phys. Rev. Lett. {\bf 86}, 5656 (2001);
Phys. Lett. {\bf B
539}, 179 (2002);
Q. R. Ahmad {\it et al.} (SNO Collaboration),
Phys. Rev. Lett. {\bf 87},
071301 (2001);
Phys. Rev. Lett. {\bf 89}, 011301 (2002);
Phys. Rev. Lett. {\bf 89}, 011302 (2002);
S. N. Ahmed {\it et al.} (SNO Collaboration),
Phys. Rev. Lett. {\bf 92}, (2004) 181301;
G. Giacomelli
{\it et al} (MACRO Collaboration),
Nucl. Phys. Proc. Suppl. {\bf 145}, 116 (2005).

\bibitem{atm}
Y.~Ashie {\it et al.}  (Super-Kamiokande Collaboration),
Phys.\ Rev.\ Lett.\  {\bf 93}   101801 (2004);
M. Ishitsuka (Super-Kamiokande Collaboration),
NOON2004, The 5th Workshop on ``Neutrino Oscillations and their origin'', 11-15 February, 2004, Tokyo, Japan;
Y.~Ashie {\it et al.}  (Super-Kamiokande Collaboration),
Phys.\ Rev.\ D {\bf 71}, 112005 (2005).

\bibitem{reac} K. Eguchi {\it et al.} (KamLAND Collaboration),
Phys. Rev. Lett. {\bf 90}, 021802 (2003); M. Apollonio {\it et al.}
(CHOOZ Collaboration), Eur.
Phys. J. C {\bf 27}, 331 (2003).


\bibitem{theta13}
A. Bandyopadhyay \etal Phys. Lett. B {\bf 608}, 115 (2005);
S.~Choubey,
Nucl.\ Phys.\ Proc.\ Suppl.\  {\bf 138}, 326 (2005).

\bibitem{nuless2beta}H. V. Klapdor-kleingrothaus, hep-ph/0512263. 

\bibitem{RT_new}
M. Gell-Mann, P. Ramond, R. Slansky, in {\em Supergravity}, ed. P. Van 
Nieuwenhuizen and D. Z. Freedman (North Holland, Amsterdam, 1979), p315;
T. Yanagida, in Proceedings of the Workshop  on the Unified Theory and the Baryon Number
in the Universe, edited by O.Sawada and A. Sugamoto (KEK-Report No. 79-18,
Tsukuba, Japan (1979)) p95;
S. Weinberg,  Phys. Rev. Lett. {\bf 43}, 1566 (1979);
R. N. Mohapatra  and  G. Senjanovic,
Phys. Rev. Lett. {\bf 44}, 912 (1980);
E. Ma, Phys. Rev. Lett. {\bf 86}, 2502 (2001).

\bibitem{branco} G. C. Branco  {\it et al}., Phys. Rev. Lett. {\bf 82}, 683 
(1999).

\end{thebibliography}
\end{document}